\documentclass[conference,10pt]{IEEEtran}
 \IEEEoverridecommandlockouts
\usepackage{cite}
\usepackage{color}
\usepackage[pdftex]{graphicx}
\graphicspath{{fig/}{jpeg/}}
\usepackage[cmex10]{amsmath}
\usepackage{amssymb}
\usepackage{algorithm}
\usepackage{algorithmic}
\usepackage{subcaption}
\usepackage[outdir=./]{epstopdf}
\usepackage{hyperref}
\usepackage{comment}
\input{mysymbol.sty}
\usepackage{needspace}

\usepackage{tikz}
\usetikzlibrary{shapes,arrows}
\usepackage{float,environ}
\usetikzlibrary{matrix} 
\usetikzlibrary{decorations.markings} 
\usetikzlibrary{calc} 
\usetikzlibrary{decorations.pathmorphing,snakes, positioning, decorations.pathreplacing}

\usetikzlibrary{fit}

\usepackage{pgfplots}
\usepackage{color}
\usepackage{multirow}

\def\BibTeX{{\rm B\kern-.05em{\sc i\kern-.025em b}\kern-.08em T\kern-.1667em\lower.7ex\hbox{E}\kern-.125emX}}
\makeatletter
\newsavebox{\measure@tikzpicture}
\NewEnviron{scaletikzpicturetowidth}[1]{%
  \def\tikz@width{#1}%
  \begin{lrbox}{\measure@tikzpicture}%
  \BODY
  \end{lrbox}%
  \pgfmathparse{#1/\wd\measure@tikzpicture}%
  \BODY
}
\makeatother

\newcommand{\QED}{\hfill\ensuremath{\blacksquare}}

\def\E{\mathbb{E}}

\def\Tr{\text{Tr}}
\def\vec2{\text{vec}}

\newtheorem{remark}{\hspace{0pt}\bf Remark}

\begin{document} 
%

\title{Control-Aware Scheduling for Low Latency \\ Wireless Systems with Deep Learning}
\author{Mark Eisen$^\dagger$ \quad Mohammad M. Rashid$^\dagger$ \quad 
			Dave Cavalcanti$^\dagger$ \quad Alejandro Ribeiro$^*$
\thanks{{Supported by Intel Science and Technology Center for Wireless Autonomous Systems. The authors are with  $(\dagger)$Intel Corporation and the $(*)$Department of Electrical and Systems Engineering, University of Pennsylvania. Email: mark.eisen@intel.com, mamun.rashid@intel.com, dave.cavalcanti@intel.com, aribeiro@seas.upenn.edu}.}}

\maketitle

\begin{abstract}
We consider the problem of scheduling transmissions over low-latency wireless communication links to control various control systems. Low-latency requirements are critical in developing wireless technology for industrial control and Tactile Internet, but are inherently challenging to meet while also maintaining reliable performance. An alternative to ultra reliable low latency communications is a framework in which reliability is adapted to control system demands. We formulate the control-aware scheduling problem as a constrained statistical optimization problem in which the optimal scheduler is a function of current control and channel states. The scheduler is parameterized with a deep neural network, and the constrained problem is solved using techniques from primal-dual learning, which have a necessary model-free property in that they do not require explicit knowledge of channels models, performance metrics, or system dynamics to execute. The resulting control-aware deep scheduler is evaluated in empirical simulations and strong performance is shown relative to other model-free heuristic scheduling methods.
\end{abstract}

\begin{IEEEkeywords}
wireless control, low-latency, deep learning, primal-dual
\end{IEEEkeywords}

\section{Introduction}

The recent advances in wireless technology and automation have given rise to efforts in integrating wireless communications in autonomous environments, particularly in industrial control and Tactile Internet settings where the scale of wired networks is proving increasingly costly \cite{ashraf2016ultra}. The analysis of control systems operating over wireless communication links is thus an integral apart in enabling these wireless industrial automation applications. However, the performance specifications of Tactile Internet applications demands the design of wireless networks that can meet both the high reliability and low latency demands of the system \cite{ ashraf2016ultra, popovski2018wireless, bennis2018ultra}. Ultra reliable low latency communications (URLLC) is inherently challenging as the physical medium of wireless communication trades off reliability and latency, making it hard to meet both demands.

One promising direction in enabling low latency communications involves specific developments in radio resource allocation, or scheduling. For low latency applications, traditional delay-aware schedulers ~\cite{wu2014analysis, lu1999fair, andrews2001providing} have been employed, in addition to more recent URLLC techniques based on various forms of diversity \cite{swamy2015cooperative, popovski2018wireless,  ashraf2018dynamic}---all of which are agnostic to the control system. However, due to the physical limitations of the wireless channel, it is often necessary to use information from the control system to make proper use of scheduling resources in meeting latency requirements. While there exist numerous ways in which control system information is incorporated into ``control-aware'' scheduling methods \cite{Cervin_event_scheduling, han2017optimal, GatsisEtal15,demirel2018deepcas,leong2018deep,silva2019optimal}, these are agnostic to latency requirements of the system. More recent work \cite{eisen2019control} looks at heuristic based scheduling methods that are both control and latency aware, but whose practical use in low latency systems is limited both by its computational complexity at every scheduling cycle and reliance on explicit knowledge of the communication model and control dynamics.

Such existing methods, however, rely on accurate system knowledge, including plant dynamics and communication network parameters. As an alternative to such traditional heuristic based scheduling methods, machine learning approaches can be incorporated into making intelligent scheduling and resource allocation decisions in wireless control systems without requiring model knowledge. The work in \cite{eisen2019learning} builds a framework for solving a generic set of resource allocation problems by interpreting resource allocation as a constrained statistical learning problem. This leads to a natural use of learning models, such as deep neural networks (DNNs), for designing schedulers. Recent advancements apply techniques from both reinforcement learning and deep learning for control-aware scheduling in simple systems \cite{demirel2018deepcas,leong2018deep,silva2019optimal} and traditional wireless systems with latency constraints \cite{elgabli2018reinforcement,sun2019unsupervised}. Learning-based scheduling policies are well suited for URLLC and control as the computational complexity at each scheduling round is very low and can furthermore be implemented model-free when system dynamics are unknown. Our contributions namely consist of
\begin{enumerate}
\item formulating a statistical learning problem for control-aware, low latency scheduling,
\item parameterizing the scheduling policy with a deep neural network (DNN), and
\item utilizing a \emph{model-free}, primal-dual learning framework to find control-aware scheduling policies.
\end{enumerate} 

This paper is organized as follows. We discuss the wireless control system in which state information is communicated to the control over a wireless channel as a switched dynamical system (Section \ref{sec_problem}). We formulate the optimal scheduling problem that minimizes a control cost under latency constraints (Section \ref{sec_opt}) and parameterize the optimal policy with a deep neural network (Section \ref{sec_param}). The constrained learning problem is solved using a so-called primal-dual learning method (Section \ref{sec_primal_dual}). We further discuss ways in which the primal-dual method can be approximated without explicit model knowledge (Section \ref{sec_model_free}). The performance of the learned control-aware scheduling method is analyzed in a numerical simulation and compared against existing baseline scheduling methods (Section \ref{sec_numerical_results}).

\section{Wireless Control Systems}\label{sec_problem}

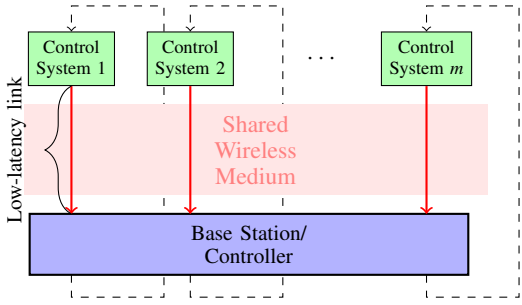
\begin{figure}
\centering
\pgfdeclarelayer{bg0}    
\pgfdeclarelayer{bg1}    
\pgfsetlayers{bg0,bg1,main}  

\tikzstyle{block} = [draw,rectangle,thick,
text height=0.2cm, text width=0.7cm, 
fill=blue!30, outer sep=0pt, inner sep=0pt]
\tikzstyle{dots} = [font = \large, minimum width=2pt]
\tikzstyle{dash_block} = [draw,rectangle,dashed,minimum height=1cm,minimum width=1cm]
\tikzstyle{smallblock} = [draw,rectangle,minimum height=0.5cm,minimum width=0.5cm,fill= green!30, font =  \scriptsize]
\tikzstyle{smallcircle} = [draw,ellipse,minimum height=0.1cm,minimum width=0.3cm,fill= yellow!40, font =  \scriptsize ]
\tikzstyle{connector} = [->]
\tikzstyle{dash_connector} = [->,thick,decorate,decoration={snake, amplitude =1pt, segment length=8pt}, magenta]
\tikzstyle{branch} = [circle,inner sep=0pt,minimum size=1mm,fill=black,draw=black]

\tikzstyle{vecArrow} = [thick, decoration={markings,mark=at position
   1 with {\arrow[semithick]{open triangle 60}}},
   double distance=1.4pt, shorten >= 5.5pt,
   preaction = {decorate},
   postaction = {draw,line width=1.4pt, white,shorten >= 4.5pt}]

\begin{tikzpicture}[scale=1, blocka/.style ={rectangle,text width=0.9cm,text height=0.6cm, outer sep=0pt}]
 \small

    \matrix(M)[ampersand replacement=\&, row sep=1.7cm, column sep=6pt] {
    
    \node[smallblock, align=center] (CS1) {Control \\ System {1}};\&\&
    \node[smallblock, align=center] (CS2) {Control \\ System {2}};\&\&\&
    \node(d1) {$\cdots$};\&
    \node[smallblock, align=center] (CSm) {Control \\ System \textit{m}};\&
    \\
    \node[blocka] (R1) {};\&\&
    \node[blocka] (R2) {};\&\&\&
    \node[blocka] (d3) {};\&
    \node[blocka] (Rm) {};\&
    \\
    };

    \node[block] (outer) [fit=(R1.north west) (d3) (Rm.south east)] {};
    
    \node[align=center, scale =0.9] at (outer.center) {Base Station/ \\Controller};
    
    \draw [->, thick, red] (CS1) -- node[left]{} (R1);
    \draw [decorate,decoration={brace,amplitude=10pt,mirror},xshift=-5]  (CS1) -- node[left]{} (R1) node [black,midway,xshift=-20.7,yshift=-0cm, rotate=90]  {\footnotesize Low-latency link};
    \draw [->, thick, red] (CS2) -- node[left]{} (R2);
    \draw [->, thick, red] (CSm) -- node[left]{} (Rm);

		\begin{pgfonlayer}{bg0}    
		\draw [->, dashed, black] (R1) |- ($(R1) + (+35pt,-20pt)$) node(down_right){} 
		-- ($(CS1) + (+35pt,+20pt)$) node(up_right){} -| (CS1);
		\end{pgfonlayer}

		\begin{pgfonlayer}{bg0}    
		\draw [->, dashed, black] (R2) |- ($(R2) + (+35pt,-20pt)$) node(down_right){} 
		-- ($(CS2) + (+35pt,+20pt)$) node(up_right){} -| (CS2);
		\end{pgfonlayer}

		\begin{pgfonlayer}{bg0}
		\draw [->, dashed, black] (Rm) |- ($(Rm) + (+35pt,-20pt)$) node(down_right){} 
		--($(CSm) + (+35pt,+20pt)$) node(up_right){} -| (CSm);
		\end{pgfonlayer}

		\begin{pgfonlayer}{bg1}
		\node(shared) [fill=red!10, fit={($(CS1.south) + (-15pt, -10pt)$) 
		($(CS2.south) + (-10pt, -10pt)$)
		($(CSm.south) + (+20pt, -10pt)$)
		($(R1.north) + (-15pt, +10pt)$)
		($(R2.north) + (-10pt, +10pt)$)
		($(Rm.north) + (+20pt, +10pt)$)
		}] {};
		\end{pgfonlayer}
		
		\node[align=center, red!50](shared_medium) at (shared.center) {Shared \\ Wireless \\ Medium};

\coordinate (FIRST NE) at (current bounding box.north east);
   \coordinate (FIRST SW) at (current bounding box.south west);

	\pgfresetboundingbox
   \useasboundingbox ($(FIRST SW) + (+30pt,0)$) rectangle (FIRST NE);

\end{tikzpicture}
\caption{A series of independent wireless control systems send state information over a shared wireless medium to a base station, where control information is fed back to the systems. The uplink transmissions (red arrow) is subject to latency constraint $t_{\max}$.}
\label{fig_wcns}
\end{figure}

We consider a series of $m$ control systems---each a wireless device or plant---operating over a shared wireless channel as shown in Figure \ref{fig_wcns}. The state of system $i$ at control cycle index $k$ is given by the variable $\bbx^k_i \in \reals^p$. At each control/scheduling cycle, the sensor measures the state $\bbx^k_i$ and transmits it over a wireless channel to a common base station (BS) that is co-located with the controller. Given the state information, the controller determines the necessary control input which is fed back to the system. This is referred to as the closed-loop configuration of the control cycle. Given the noisy nature of the wireless channel, there is the potential for the communications packet containing the state information to be dropped, resulting in an open-loop configuration of the control cycle. We may model the linear dynamics of the wireless control system for system $i$ as
\begin{equation}\label{eq:system}
	\bbx_i^{k+1} = \left\{ \begin{array}{ll} \hbA_i \bbx^k_i + \bbw^k &\text{if packet received} \\ \mathring{\bbA}_i \bbx^k_i + \bbw^k & \text{otherwise} \end{array} \right.,
\end{equation}
where $\hbA_i \in \reals^{p\times p}$ is the closed loop gain, $\mathring{\bbA}_i \in \reals^{p\times p}$ is the open loop gain, and $\bbw^k \in \reals^p$ is zero-mean i.i.d. disturbance process with covariance $\bbW$. The closed loop and open loop gains may reflect, e.g., controlled dynamics using accurate and estimated state information, respectively. We assume that the closed loop gains are preferable to the open loop gain, i.e. $\lambda_{\max}(\hbA_i) < \lambda_{\max}(\mathring{\bbA}_i)$. Further note this model restricts its attention to wireless connections in uplink of the control loop, while downlink is assumed to occur over an ideal channel---i.e. no packet drops.
\begin{figure}
\centering
\resizebox{7cm}{3.2cm}{%
\begin{tikzpicture}
\draw[black,very thick] (-4,-1) rectangle (1,0);
\draw[black,very thick] (-4,0) rectangle (0.75,1);
\draw[black,very thick] (-4,1) rectangle (1.2,2);
\draw [step=.3, thick, draw= black, fill=black!10!green!30] (-4,-1) rectangle  (0,0);
\draw [step=.3, thick, draw= black, fill=black!10!blue!30] (0,-1) rectangle  (1,0);
\draw [step=.3, thick, draw= black, fill=black!10!red!30] (-4,0) rectangle  (-2,1);
\draw [step=.3, thick, draw= black, fill=black!10!blue!30] (-2,0) rectangle  (-1.2,1);
\draw [step=.3, thick, draw= black, fill=black!10!green!30] (-1.2,0) rectangle  (0.75,1);
\draw [step=.3, thick, draw= black, fill=black!10!yellow!30] (-4,1) rectangle  (1.2,2);
\node[align=center, scale =0.9] at (-1.5,-1.5) {Transmission time};
\node[align=center, scale =0.9] at (-5,-.5) {Channel 3};
\node[align=center, scale =0.9] at (-5,.5) {Channel 2};
\node[align=center, scale =0.9] at (-5,1.5) {Channel 1};
\end{tikzpicture}
}
  
\label{fig_multi}
\end{figure}

Given this dynamical model of the wireless control systems, the communications goal is to allocate radio resources among the various systems to maintain strong performance across all the systems. To do so, we present a generic frequency and time division multiplexing scheduling architecture with which the BS allocates scheduling resources to the systems. A scheduling window occupies the uplink of a single cycle in the control loop in which each system has a single packet containing state information to transmit. For URLLC systems, the total length of this scheduling window is subject to a tight low-latency bound $t_{\max}$. 

We assume that transmissions are scheduled by the BS across $n$ available channels occupying different (possibly non-consecutive) frequency bands. Each channel is subject to continuous time division multiple access (TDMA), meaning that multiple transmissions in the same channel will occur in sequence. For full generality, we assume that a single device may be scheduled in multiple channels in a single cycle to add redundancy and improve chance of success. Denote by $\bbvarsigma_i \in \{0,1\}^n$ a binary vector whose $j$th element $\varsigma_{i,j}$ is 1 if the $i$th device transmits in the $j$th channel, and 0 otherwise. Further denote for each device a data rate selection $\mu_i \in [\mu_{\min}, \mu_{\max}]$. These two scheduling parameters together define the scheduling decision made for the $i$th system. An illustration of $m=4$ users making multiple transmission across $n=3$ channels is shown in Figure \ref{fig_multi}.

The achieved communications performance by a given scheduling decision can be formulated as follows. We first define $\bbh^k_{i} \in \reals^n_+$ to be the set of fading channel states experienced by device $i$ at cycle $k$, where the $j$ element $h^k_{i,j}$ is the fading channel gain in channel $j$. We assume that these channel conditions do not change over the course of a scheduling window. In any given channel with fading state $\bbh$, we define a function $q(h,\mu)$ that returns the packet delivery rate (PDR), or the probability of successful transmission of the packet, when transmitting with data rate $\mu$. Likewise, we define a function $\tau(\mu)$ that returns the transmission time to transmit a packet of fixed length with data rate $\mu$. These two functions play a critical role in designing low-latency wireless control systems, as they allow us to explore the trade-off between PDR and transmission time and the resulting effect on control system performance. We may consider that the functions $q(h, \mu)$ and $\tau(\mu)$ both get smaller as we increase data rate $\mu$, i.e.
\begin{equation}
\mu' > \mu \implies q(h, \mu) \leq q(h, \mu'), \quad \tau(\mu') \leq \tau(\mu).
\end{equation}
Thus, by increasing the data rate we may reduce the transmission time to satisfy latency constraints, but at the cost of control system performance, as illustrated by the switched dynamics in \eqref{eq:system}.

\begin{remark}
The communication architecture utilized here has a generic form that assumes both continuous time division and simultaneous transmission in independent, unsynchronized channels. We present the architecture in this form both for the  purposes of a more tractable mathematical model as well as its generalization of the architectures used in, e.g., Bluetooth or centralized scheduled WiFi. Note that common OFDMA architectures, such as 5G \cite{agiwal2016next} and next-generation WiFi IEEE 802.11ax \cite{liu2014ieee}, do not conform precisely to this architecture although it can be adapted as such with slight modifications. We leave the consideration of a synchronized, OFDMA architecture as a point of future work.
\end{remark}

\subsection{Optimal scheduling design}\label{sec_opt}

We are interested in designing scheduling policies that optimize control performance, subject to the strict low latency constraints of the system. To do so, we first formulate the global control-based performance given a scheduling decision. Collect in the matrix $\bbSigma \in \{0,1\}^{n \times m}$ all of the channel transmission vectors $\bbvarsigma_i$ for $i=1,\hdots,m$ and collect in the vector $\bbmu \in [\mu_{\min},\mu_{\max}]^m$ the data rates $\mu_i$ for $i=1,\hdots,m$. Given that a device may transmit in multiple channels within a single scheduling cycle, the probability of successful transmission can be given as the probability that the transmission was successful in at least one channel, i.e.
\begin{equation}\label{eq_psr}
\tdq(\bbh_i, \bbvarsigma_i, \mu_i) := 1 - \prod_{j=1}^n \left(1 - \varsigma_{i,j} q(h_{i,j}, \mu_i)\right). 
\end{equation}

The total delivery rate in \eqref{eq_psr} can be viewed as the probability of receiving the packet and experiencing the closed loop dynamics in \eqref{eq:system}. Now, to evaluate the performance of a given system at a particular state $\bbx$, define a quadratic Lyapunov function $L_i(\bbx) := \bbx^T \bbP_i \bbx$ with some positive definite matrix $\bbP_i \in \reals^{p \times p}$. Such a function can be used to evaluate performance or stability of the control system. Because the control system evolves in a random manner, the cost of a given scheduling decision $\{\bbvarsigma_i, \mu_i\}$ for the $i$th system can be formulated as the \emph{expected future Lyapunov cost} under such a schedule. As the probability of closing the loop in \eqref{eq:system} is given by $\tdq(\bbh^k_i, \bbvarsigma_i, \mu_i)$, we may write this expected future cost as 
\begin{align}\label{eq_ex_cost}
J_i(\bbx_i, \bbh_i, \bbvarsigma_i, \mu_i) :=& \quad \E \left[ L_i(\bbx^{k+1}_i) \mid \bbx^k_i = \bbx_i, \bbh^k_i = \bbh_i \right] \\
=&  \quad  \tdq(\bbh_i, \bbvarsigma_i, \mu_i) (\hbA_i \bbx_i)^T \bbP_i (\hbA_i \bbx_i) \quad +  \nonumber   \\
 &(1- \tdq(\bbh_i, \bbvarsigma_i, \mu_i))  (\mathring{\bbA}_i \bbxi)^T \bbP_i (\mathring{\bbA}_i \bbx_i) \nonumber \\
 &+  \quad \Tr(\bbP_i \bbW_i). \nonumber
\end{align}
Observe that the local control cost for the $i$th system $J_i(\bbx^k_i, \bbh^k_i, \bbvarsigma_i, \mu_i)$ is a function of both the system \emph{states}---the fading channel $\bbh^k_i$ and control state $\bbx^k_i$---and the scheduler \emph{actions}---channel selection $\bbvarsigma_i$ and data rate $\mu_i$. The objective is to choose the actions $\bbvarsigma_i$ and $\mu_i$ that minimizes the cost relative to states $\bbh^k_i$ and $\bbx^k_i$. 

In addition to minimizing a control cost, we must make scheduling decisions that respect the low-latency requirements of the system. To formulate this constraint, consider the \emph{total} time of a global scheduling decision $\bbSigma, \bbmu$ of channel $j$ as the sum of all active transmissions, i.e.
\begin{equation} \label{eq_total_time}
\tilde{\tau}_j(\bbSigma, \bbmu) :=   \sum_{i=1}^m \varsigma_{i,j} \tau(\mu_i).
\end{equation}

Combining all the local costs for systems $i=1,\hdots,m$ in \eqref{eq_ex_cost} with the a constraint on the latency costs for all channels $j=1,\hdots,n$ in \eqref{eq_total_time}, we may define the optimal scheduling design problem. Because we are interested in long-term, or average, performance across random channels and control states, we optimize with respect to expected costs and probabilistic constraints. Collect all channel vectors $\bbh_i$ in a matrix $\bbH \in \reals_+^{n \times m}$ and states $\bbx_i$ in a matrix $\bbX \in \reals^{p \times n}$. Consider a scheduling policy $\bbp(\bbH, \bbX) := \{ \bbSigma, \bbmu\}$ that, given a set of channel states $\bbH$ and control states $\bbX$, returns a schedule defined by the channel selection matrix $\bbSigma$ and data rate selection vector $\bbmu$. The optimal low-latency constrained scheduling policy for the wireless control systems is the one which solves the program
\begin{alignat}{2} \label{eq_problem}
   J^* := &  \min_{\bbp(\bbH, \bbX)}  \E_{\bbH,\bbX} \left[ \sum_{i=1}^m J_i(\bbx_i,\bbh_i, \bbvarsigma_i, \mu_i) \right],             \\
        &  \st           \quad           \mathbb{P}_{\bbH,\bbX} \left( \tilde{\tau}_j(\bbSigma, \bbmu) \leq t_{\max} \right)   \geq 1-  \delta \quad j=1\hdots,n,   \nonumber \\
        &     \                       \bbp(\bbH,\bbX) := \{ \bbSigma \in  \{0,1\}^{n \times m}, \bbmu \in [\mu_{\min},\mu_{\max}]^m \}.   \nonumber%
\end{alignat}
In \eqref{eq_problem}, we minimize the average cost over the distribution of channel and control states, subject to the condition that the probability of violating the latency constraint over the distribution of states is less than some small value $\delta$. Because each channel's transmission time varies, we impose this constraint independently for \emph{each channel}. The above scheduling problem can be viewed as a constrained statistical learning problem---a connection made for a more generic class of resource allocation problems in \cite{eisen2019learning}. While such a problem characterizes the optimal scheduling decision for the latency-constraint wireless control system, finding solutions to such a problem is a significant challenge. This is due to a number of complexities in \eqref{eq_problem}, namely: (i) it requires functional optimization, (ii) it contains explicit constraints, and (iii) we typically do not have analytic forms for the functions and distributions in \eqref{eq_problem}. The first of these complexities can be resolved using a standard technique in statistical learning, discussed next in Section \ref{sec_param}. The latter two of these complexities are discussed and resolved later in Sections \ref{sec_primal_dual} and \ref{sec_model_free}, respectively.

\subsection{Deep learning parameterization}\label{sec_param}

The scheduling problem in \eqref{eq_problem} is computationally challenging because it requires finding a policy---or \emph{function}---$\bbp(\bbH,\bbX)$. In statistical learning, or regression, problems the regression function is replaced by some given parameterization $\bbphi(\bbH,\bbX,\bbtheta)$ that is defined with some finite dimensional parameter $\bbtheta \in \reals^q$. There exist a wide variety of choices of this parameterization, but in modern machine learning problems the \emph{deep neural network (DNN)} is commonly employed. This is due to the fact the DNN can be shown both empirically and analytically to contain strong representative power and generalization ability, meaning that it can approximate almost any function well. A DNN is defined as a composition of $L$ layers, each of which consisting of a linear operation followed by a point-wise nonlinearity---also known as an activation function.  More specifically, the layer $l$ is defined by the linear operation $\bbW_l \in \reals^{q_{l-1} \times q_l}$ followed by a non-linear activation function $\mathbb{\sigma}_{l}: \reals^{q_l} \rightarrow \reals^{q_l}$. Common choices of activation functions $\mathbb{\sigma}_{l}$ include a sigmoid function or a rectifier function (commonly referred to as ReLu). Given an input from the $l-1$ layer $\bbw_{l-1} \in \reals^{q_{l-1}}$, the resulting output $\bbw_{l} \in \reals^{q_l}$ is then computed as $\bbw_l := \mathbb{\sigma}_{l}(\bbW_l \bbw_{l-1})$.  The full DNN-parameterization of the scheduling policy is then defined as an $L$-layer DNN whose input at the initial layer is the concatenation of states $\bbw_0 := [\vec2(\bbH); \vec2(\bbX)]$, i.e.
\begin{equation}\label{eq_policy_dnn}
\bbphi(\bbH, \bbX,\bbtheta) := \mathbb{\sigma}_{L}(\bbW_{L} (\mathbb{\sigma}_{L-1}(\bbW_{L-1}(\hdots(\mathbb{\sigma}_{1}(\bbW_{1}\bbw_0)))))).
\end{equation}

The parameter vector $\bbtheta \in \reals^q$ that defines the DNN is then the entries of $\{ \bbW_l \}_{l=1}^L$ with $q = \sum_{l=1}^{L-1} q_l q_{l+1} $. Further note that we can easily construct an activation function at the final layer $\mathbb{\sigma}_L$---or the \emph{output layer}---such that the outputs $\bbphi(\bbH,\bbX,\bbtheta)$ are in the space $\{0,1\}^{n \times m} \times [\mu_{\min},\mu_{\max}]$ that contains possible schedules. With this DNN parameterization, the control-aware scheduling problem can be rewritten as
\begin{alignat}{2} \label{eq_param_problem}
   J_{\bbphi}^* := &  \min_{\bbtheta \in \reals^q}  \E_{\bbH,\bbX} \left[ \sum_{i=1}^m J_i(\bbx_i,\bbh_i, \bbsigma_i, \mu_i) \right],             \\
        &  \st           \quad           \mathbb{P}_{\bbH,\bbX} \left( \tilde{\tau}_j(\bbSigma, \bbmu) \leq t_{\max} \right)   \geq 1-  \delta \qquad \forall j,   \nonumber \\
        &     \                       \bbphi(\bbH,\bbX,\bbtheta) := \{ \bbSigma \in  \{0,1\}^{n \times m}, \bbmu \in [\mu_{\min},\mu_{\max}]^m \}.   \nonumber%
\end{alignat}
Observe in \eqref{eq_param_problem} that the optimization is performed over $\bbtheta$ rather than the scheduling policy directly. In other words, we look for the interlayer weights that define a DNN that minimizers the total control cost while satisfying the latency constraints. We proceed then to discuss a learning method that can find solutions to the constrained optimization problem in \eqref{eq_param_problem}.

\section{Primal-Dual Learning}\label{sec_primal_dual}

Finding the DNN layer weights $\bbtheta$ that provide good solutions to \eqref{eq_param_problem} requires the solving of a constraint learning problem. The standard approach of gradient-based optimization methods cannot be applied directly here due to the presence of the latency constraints. To proceed then, we must formulate an unconstrained problem that captures the form of \eqref{eq_param_problem}. A naive penalty-based reformulation will introduce a similar but fundamentally different problem, so we thus opt for constructing a Lagrangian dual problem. For notational convenience, moving forward we employ the following shorthands for the state variables, aggregate Lyapunov function, latency constraint functions, respectively:
\begin{align}
\bbw &:= [\vec2(\bbH); \vec2(\bbX)], \\
f(\bbphi(\bbw,\bbtheta), \bbw) :&= \sum_{i=1}^m J_i(\bbx_i, \bbh_i, \bbvarsigma_i, \mu_i), \\
g_j(\bbphi(\bbw,\bbtheta),\bbw) &:=  \mathbb{I}\left[ \tdtau_j(\bbSigma,\bbmu) \leq t_{\max}\right] - (1 - \delta)
\end{align}
We introduce the nonnegative dual variables $\bblambda \in \reals_+^n$ associated with the vector of constraint functions $\bbg(\bbp(\bbw,\bbtheta),\bbw) := [g_1(\cdot); \hdots; g_n(\cdot)]$, and form the Lagrangian as
\begin{align}\label{eq_param_lagrangian}
   \ccalL(\bbtheta,\bblambda) &:=   \E_{\bbw} \left[ f(\bbphi(\bbw,\bbtheta), \bbw) - \bblambda^T \bbg(\bbphi(\bbw,\bbtheta),\bbw) \right].
\end{align}
The Lagrangian in \eqref{eq_param_lagrangian} penalizes constraint violation through the second term. Note, however, that the penalty is scaled by the dual parameter $\bblambda$. The so-called Lagrangian dual problem is one in which both the primal variable $\bbtheta$ is simultaneously minimized while the dual parameter $\bblambda$ is maximized. Such a problem can be written with the saddle point formulation
\begin{align}\label{eq_param_dual0}
   D_{\bbphi}^* &:= \max_{\bblambda \geq \bb0} \min_{\bbtheta}\ccalL(\bbtheta,\bblambda).
\end{align}
The dual optimum $D_{\bbphi}^*$ is the best approximation of the form in \eqref{eq_param_lagrangian} we can have of $J_{\bbphi}^*$. In fact, under some standard assumptions on the problem and assuming a sufficiently dense DNN architecture, we can formally bound the difference between $D_{\bbphi}^*$ and $J^*$ to be proportional to the approximation capacity of the DNN $\bbphi(\bbH,\bbX,\bbtheta)$---see \cite{eisen2019learning} for details on this result. Thus, we may say that, up to some approximation, solving the unconstrained problem in \eqref{eq_param_dual0} is equivalent to solving the constrained problem in \eqref{eq_param_problem}.

With the unconstrained saddle point problem in \eqref{eq_param_dual0}, we may perform standard gradient-based optimization methods to obtain solutions. The max-min structure necessitates the use of a \emph{primal-dual} learning method, in which we iteratively update both the primal and dual variable in \eqref{eq_param_lagrangian} to find a local stationary point of the KKT conditions of \eqref{eq_param_problem}. Consider a learning iteration index $t=0,1,\hdots$ over which we define a sequence of primal variables $\{\bbtheta_t\}$ and dual variables $\{\bblambda_t\}$. At index $t$, we determine the value of next primal iterate $\bbx_{t+1}$ by adding to the current iterates the corresponding partial gradients of the Lagrangian in \eqref{eq_param_lagrangian} $\nabla_{\bbtheta} \ccalL$, i.e., 
\begin{align}
\bbtheta_{t+1} \! &= \bbtheta_t \! - \! \alpha_t  \nabla_{\bbtheta}\E_{\bbw}  \left[ f(\bbphi(\bbw,\bbtheta_t),\bbw) \! - \! \bblambda_t^T \bbg(\bbphi(\bbw,\bbtheta_t),\bbw) \right] \label{eq_pd_update1},
\end{align}
where we introduce $\alpha_t >0$ as a scalar step size. We subsequently perform a corresponding partial gradient update to compute the dual iterate $\bblambda_{t+1}$, i.e.
\begin{align}
\bblambda_{t+1} &= \left[ \bblambda_t - \beta_t \E_{\bbw} \bbg(\bbphi(\bbw,\bbtheta_{t+1}),\bbw) \right]_+,\label{eq_pd_update2}
\end{align}
with associated step size $\beta_t >0$. Observe in \eqref{eq_pd_update2} that we additionally project onto the positive orthant to maintain the nonnegative constraint on $\bblambda$. The gradient primal-dual updates in \eqref{eq_pd_update1} and \eqref{eq_pd_update2} successively move the primal and dual variables towards maximum and minimum points of the Lagrangian function, respectively.

\subsection{Model-free updates}\label{sec_model_free}

The updates in \eqref{eq_pd_update1}-\eqref{eq_pd_update2} cannot, in general, be applied exactly. To see this, observe that computing the gradients in \eqref{eq_pd_update1} requires computing the gradient of $J_i(\cdot)$---which depends on PDR function $\tdq(\cdot)$ and system dynamics---and the gradient of an indicator of  transmission length function $\tdtau(\cdot)$. In practical systems, we do not typically have easily available analytic forms for these functions to take gradients. Furthermore, both the updates in \eqref{eq_pd_update1} and \eqref{eq_pd_update2} require to take the expectation over the distribution of states $\bbx$ and $\bbh$. These, too, are often unknown in practice. However, there exist standard ways of approximating the updates with stochastic, \emph{model-free} updates that do not require such knowledge. Most popular among these is the policy gradient approximation \cite{sutton2000policy}. 

To compute a policy gradient update, we consider the scheduling parameters $\bbSigma$ and $\bbmu$ are drawn stochastically from a distribution with given form $\pi_{\bbphi(\bbw,\bbtheta)}$ whose parameters are given by the output of the DNN $\bbphi(\bbw,\bbtheta)$---e.g. the mean and variance of a normal distribution. Using such a stochastic policy, it can be shown that an unbiased estimators of the gradients in \eqref{eq_pd_update1} and \eqref{eq_pd_update2} can be formed as,
\begin{align}\label{eq_policy_gradient_s}
\widehat{\nabla_{\bbtheta}} \mathbb{E}_{\bbw} f(\bbphi(\bbw,\bbtheta),\bbw) \!&=\!  f( \hbp_{\bbtheta},\hbw) \nabla_{\bbtheta} \log \pi_{\bbphi(\hbw,\bbtheta)}(\hbp_{\bbtheta} ) \\
\widehat{\nabla_{\bbtheta}} \mathbb{E}_{\bbw} \bbg(\bbphi(\bbw,\bbtheta),\bbw)\! &= \! \bbg( \hbp_{\bbtheta},\hbw) \nabla_{\bbtheta} \log \pi_{\bbphi(\hbw,\bbtheta)}(\hbp_{\bbtheta} )^T \\
\widehat{ \mathbb{E}_{\bbw}} \bbg(\bbphi(\bbw,\bbtheta),\bbw) &= \bbg(\hbp_{\bbtheta},\hbw), \label{eq_pg_3}
\end{align}
where $\hbw$ is a sampled state and $\hbp_{\bbtheta}$ is a sample drawn from the distribution $\pi_{\bbphi(\hbw,\bbtheta)}$. In practice, we may reduce the variance of these unbiased estimates by taking $B$ samples and averaging. Note that the updates here only require taking gradients of the log likelihoods rather than of the functions themselves. This implies we can perform the learning process without explicitly knowing, e.g., system dynamics, performance metrics, state distributions. Thus, we can replace the updates in \eqref{eq_pd_update1} and \eqref{eq_pd_update2} with their model free counterparts by substituting the gradient estimates in \eqref{eq_policy_gradient_s}-\eqref{eq_pg_3}. The complete primal-dual learning algorithm is summarized in Algorithm \ref{alg:learning}. We conclude with a brief remark on state sampling.
{\linespread{1.3}
\begin{algorithm}[h] \begin{algorithmic}[1]
\STATE \textbf{Parameters:} Policy model $\bbphi(\bbh,\bbtheta)$ and distribution $\pi_{\bbh,\bbtheta}$ 
\STATE \textbf{Input:} Initial states $\bbtheta_0, \bblambda_0$
\FOR [main loop]{$t = 0,1,2,\hdots$}
      \STATE Draw samples $\{  \hbtheta, \hbh \}$, or in batches of size $B$
      \STATE Compute policy gradients  [ c.f. \eqref{eq_policy_gradient_s}-\eqref{eq_pg_3}]
      \STATE Update primal and dual variables  \ 
      \footnotesize
         \begin{align}
	\mkern-36mu \bbtheta_{t+1} \! &= \bbtheta_t \! - \! \alpha_t  \widehat{\nabla_{\bbtheta}}\E_{\bbw}  \left[ f(\bbphi(\bbw,\bbtheta_t),\bbw) \! - \! \bblambda_t^T \bbg(\bbphi(\bbw,\bbtheta_t),\bbw) \right] , [cf. \eqref{eq_pd_update1}]\nonumber \\
	\mkern-36mu \bblambda_{t+1} &= \left[ \bblambda_t - \beta_t \widehat{\E_{\bbw}} \bbg(\bbphi(\bbw,\bbtheta_{t+1}),\bbw) \right]_+ \!\![cf. \eqref{eq_pd_update2}]\	\nonumber 
\end{align}
\ENDFOR

\end{algorithmic}
\caption{Model-Free Primal-Dual Learning}\label{alg:learning} \end{algorithm}}

\begin{remark}\label{remark}\normalfont
In the gradient estimations in, e.g. \eqref{eq_policy_gradient_s}, we sample both the control states $\bbx$ and channel states $\bbh$. This assumes that such samples can be drawn i.i.d. While this may generally be true for the channel states $\bbh$, it will not be generally be true for the control states $\bbx$ in practice, due to the fact that the states evolve based on the switched dynamics in \eqref{eq:system}, which itself depends on the scheduling actions taken. A more precise way to model the statistics of the control states would be with a Markov decision process (MDP). The generalization of the presented techniques for this setting make up what is known as \emph{reinforcement learning} algorithms. In this work, we nonetheless assume that $\bbx$ can also be drawn i.i.d. from an approximate distribution and leave the full MDP formulation as the study of future work.
\end{remark}

 \begin{figure}[t]
\centering
\includegraphics[width=.45\textwidth, height=.22\textheight]{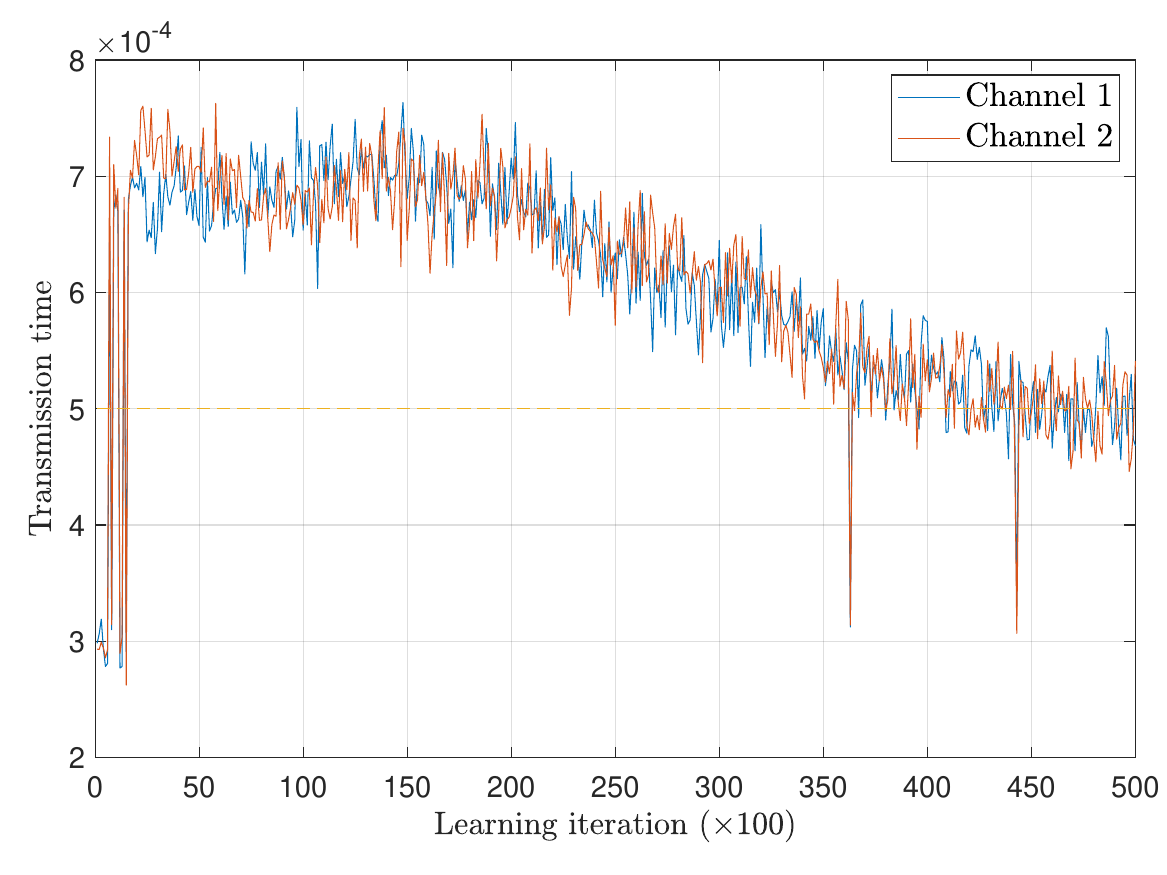}
\caption{Convergence of transmission time for a low-latency, control aware scheduling policy over the learning process. The DNN parameterized scheduling policy obtains feasible latency-contained schedules ($t_{\max} = 5 \times 10^{-4}$ shown in dashed red line) on both channels. }\label{fig_simple_results0}
\end{figure}

 \begin{figure*}[t]
\centering
\includegraphics[width=.45\textwidth, height=.22\textheight]{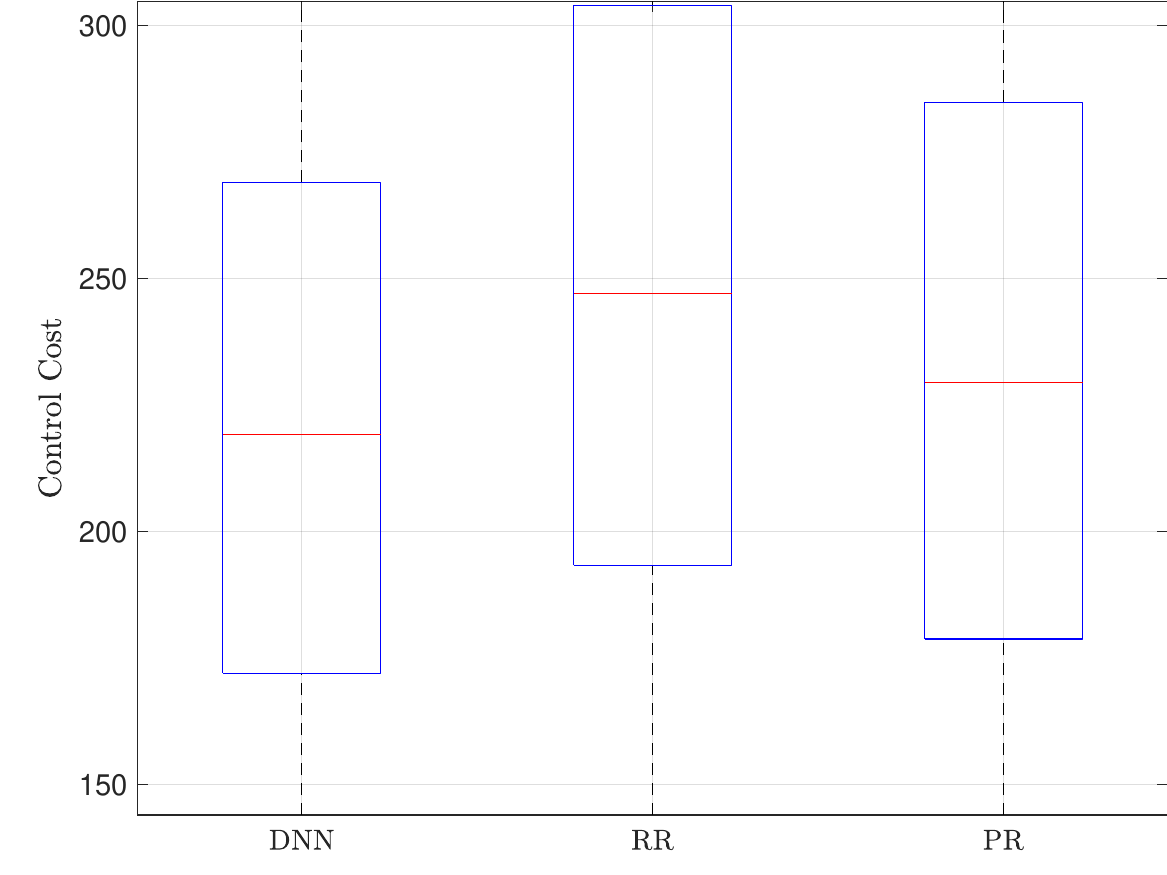} \quad
\includegraphics[width=.45\textwidth, height=.24\textheight]{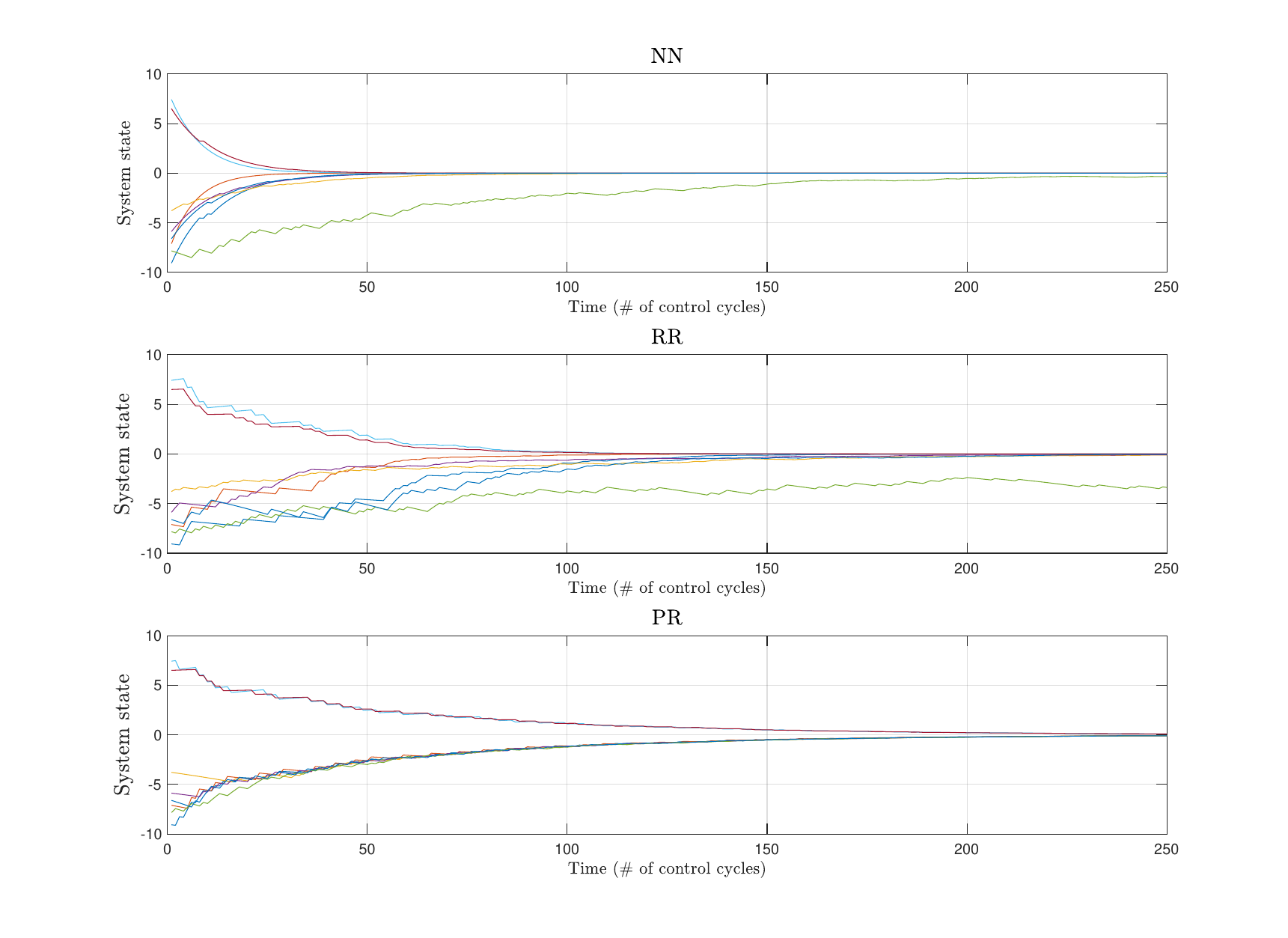}
\caption{(Left) A boxplot of medians and quartiles of control costs obtained by proposed DNN-based scheduling and model-free heuristics. In terms of the quadratic system cost, the DNN outperforms both baselines. In (Right) we simulate the control system using the learned scheduler and two baseline model-free heuristic scheduler. The NN policy stabilizes most of the systems faster than RR and PR.}\label{fig_simple_results}
\end{figure*}

 \section{Simulation Results}\label{sec_numerical_results}

We perform a series of simulations on latency-constrained wireless control systems to evaluate the performance the learning method in and the resulting control-aware scheduling policies. We generate a series $m=9$ systems with closed-loop gains $\hbA_i \sim \text{Uniform}(0.85,0.95)$ and open-loop gains $\mathring{\bbA}_i \sim \text{Uniform}(1.01,1.2)$. The variance for all system noise $\bbw_i$ is set to be $W=1$. All such systems send their state information over a shared wireless channel with $n=2$ independent channels with a total latency constraint of $t_{\max} = 0.5$ ms. A latency bound of this order is typical of industrial control systems such as printing machines and presses \cite{ashraf2016ultra}. We further assume that the states of the systems are confined to the box $[-10,10]$. In simulations, we utilize a DNN with 2 layers of size 2000 and 1000, with ReLU activation functions. For the policy distribution $\pi_{\bbphi(\bbw,\bbtheta)}$, we utilize a Beta distribution scaled between $ [1.6,9.0]$ to select $\bbmu$ and a Bernoulli distribution to select $\bbSigma$.

With the scheduling architecture given in Figure \ref{fig_multi} for 2 channels and 9 systems, at each control scheduling interval each system is given a data rate $\mu_i$ and a set of channels to transmit on. In our simulations, we use the modulation and coding schemes (MCS) of the next-generation IEEE 802.11ax Wi-Fi protocol as a representative architecture for data rate selection and packet error rate computation. As such, the continuous data rates $\mu_i$ are selected in an interval of $[1.6, 10]$ and rounded down to the nearest discrete MCS selection given in 802.11ax---see \cite{liu2014ieee} for details on the MCS tables given in this protocol. The corresponding transmission time $\tau(\mu)$ is then calculated assuming a fixed packet size of 100 bytes and the packet delivery rate $q(h,\mu)$ is computed using the associated AWGN error curve (scaled by the effective SNR given channel conditions).

In Figure \ref{fig_simple_results0} we show the training process for the control-aware scheduler using the primal-dual scheduler given in Algorithm \ref{alg:learning}. In particular, we show the transmission time utilized over the 2 channels by the NN-based scheduling policy over the course of $50,000$ learning iterations. As can be seen, the policies converge to scheduling decisions that respect latency requirements for both channels after $50,000$ iterations, showcasing the capability of a neural network to learn latency restrictive policies.

We proceed to compare the performance of the learned policy in terms of the control metric in \eqref{eq_ex_cost} against other scheduling heuristics. We compares against a standard, control-agnostic round-robin scheduling policy (RR) and a control-aware priority ranking (PR) heuristic in which transmissions are prioritized for systems with largest states. These methods are chosen as they can be implemented model-free, to make a reasonable comparison against the model-free DNN, and are commonly used in modern practical systems. We point out that both of the scheduling policies used fixed PDRs of 0.95 to determine data rate selection. In the left image of Figure \ref{fig_simple_results} we show a box plot of the quadratic control costs obtained by each of these methods over 1000 different randomly drawn plant and channel states. It can be observed that, in terms of this cost, the DNN outperforms both baselines. Thus, when considering a specific control-aware cost to optimize, designing scheduling algorithms directly with respect to this cost can benefit the performance of the system.

Alternatively to the quadratic cost shown in the left image, we may observe the end system performance of each of the scheduling methods by looking at the state evolution of each of the 9 plants using the respective schedulers. In the right image of Figure \ref{fig_simple_results} we show evolution of the 9 systems under each method. It can be observed that, while all systems stabilize using each of the three schedulers, the DNN is overall able to draw the plant states to zero faster than the other methods, with one exception. Together, the results in both figures of Figure \ref{fig_simple_results} demonstrate an improved performance relative to existing baselines. This can be attributed to the fact that DNN has been model-free trained to adapt to both changing channel conditions \emph{and} the individual dynamics and states of each of the systems, which allows it to make more efficient scheduling policies with regards to the varying system dynamics and latency constraints. In these results we observe that it is indeed advantageous to incorporate control system knowledge in the scheduling decision to promote good performance.

 
 \section{Conclusion}
We consider the setting of scheduling for low-latency wireless control systems. To handle the challenge of achieving high reliability performance with limited scheduling resources, we formulate a control-aware scheduling problem in which reliability is adapted to control and channel states. This problem takes the form of a constrained statistical learning problem, in which solutions can be found by parameterized the scheduling policy with a deep neural network and finding optimal weights with a primal-dual learning algorithm that can be implemented without system or dynamical models. Numerical simulations showcase DNN-based scheduling policies that outperform baseline scheduling procedures.

\urlstyle{same}
\bibliographystyle{IEEEbib}
\bibliography{wireless_ll_control,scheduling_control}

\begin{thebibliography}{10}

\bibitem{ashraf2016ultra}
Shehzad~A Ashraf, Ismet Aktas, Erik Eriksson, Ke~Wang Helmersson, and Junaid
  Ansari,
\newblock ``Ultra-reliable and low-latency communication for wireless factory
  automation: From {LTE} to {5G},''
\newblock in {\em Emerging Technologies and Factory Automation (ETFA), 2016
  IEEE 21st International Conference on}. IEEE, 2016, pp. 1--8.

\bibitem{popovski2018wireless}
Petar Popovski, Jimmy~J Nielsen, Cedomir Stefanovic, Elisabeth de~Carvalho,
  Erik Strom, Kasper~F Trillingsgaard, Alexandru-Sabin Bana, Dong~Min Kim,
  Radoslaw Kotaba, Jihong Park, et~al.,
\newblock ``Wireless access for ultra-reliable low-latency communication:
  Principles and building blocks,''
\newblock {\em IEEE Network}, vol. 32, no. 2, pp. 16--23, 2018.

\bibitem{bennis2018ultra}
Mehdi Bennis, M{\'e}rouane Debbah, and H~Vincent Poor,
\newblock ``Ultra-reliable and low-latency wireless communication: Tail, risk
  and scale,''
\newblock {\em arXiv preprint arXiv:1801.01270}, 2018.

\bibitem{wu2014analysis}
Chengjie Wu, Mo~Sha, Dolvara Gunatilaka, Abusayeed Saifullah, Chenyang Lu, and
  Yixin Chen,
\newblock ``Analysis of edf scheduling for wireless sensor-actuator networks,''
\newblock in {\em Quality of Service (IWQoS), 2014 IEEE 22nd International
  Symposium of}. IEEE, 2014, pp. 31--40.

\bibitem{lu1999fair}
Songwu Lu, Vaduvur Bharghavan, and Rayadurgam Srikant,
\newblock ``Fair scheduling in wireless packet networks,''
\newblock {\em IEEE/ACM Transactions on networking}, vol. 7, no. 4, pp.
  473--489, 1999.

\bibitem{andrews2001providing}
Matthew Andrews, Krishnan Kumaran, Kavita Ramanan, Alexander Stolyar, Phil
  Whiting, and Rajiv Vijayakumar,
\newblock ``Providing quality of service over a shared wireless link,''
\newblock {\em IEEE Communications magazine}, vol. 39, no. 2, pp. 150--154,
  2001.

\bibitem{swamy2015cooperative}
Vasuki~Narasimha Swamy, Sahaana Suri, Paul Rigge, Matthew Weiner, Gireeja
  Ranade, Anant Sahai, and Borivoje Nikoli{\'c},
\newblock ``Cooperative communication for high-reliability low-latency wireless
  control,''
\newblock in {\em Communications (ICC), 2015 IEEE International Conference on}.
  IEEE, 2015, pp. 4380--4386.

\bibitem{ashraf2018dynamic}
Muhammad~Ikram Ashraf, Chen-Feng Liu, Mehdi Bennis, Walid Saad, and Choong~Seon
  Hong,
\newblock ``Dynamic resource allocation for optimized latency and reliability
  in vehicular networks,''
\newblock {\em IEEE Access}, vol. 6, pp. 63843--63858, 2018.

\bibitem{Cervin_event_scheduling}
Anton Cervin and Toivo Henningsson,
\newblock ``Scheduling of event-triggered controllers on a shared network,''
\newblock in {\em Proc. of the 47th IEEE Conf. on Dec. and Control (CDC)},
  2008, pp. 3601--3606.

\bibitem{han2017optimal}
Duo Han, Junfeng Wu, Huanshui Zhang, and Ling Shi,
\newblock ``Optimal sensor scheduling for multiple linear dynamical systems,''
\newblock {\em Automatica}, vol. 75, pp. 260--270, 2017.

\bibitem{GatsisEtal15}
Konstantinos Gatsis, Miroslav Pajic, Alejandro Ribeiro, and George~J. Pappas,
\newblock ``Opportunistic control over shared wireless channels,''
\newblock {\em IEEE Transactions on Automatic Control}, vol. 60, no. 12, pp.
  3140--3155, December 2015.

\bibitem{demirel2018deepcas}
Burak Demirel, Arunselvan Ramaswamy, Daniel~E Quevedo, and Holger Karl,
\newblock ``Deepcas: A deep reinforcement learning algorithm for control-aware
  scheduling,''
\newblock {\em IEEE Control Systems Letters}, vol. 2, no. 4, pp. 737--742,
  2018.

\bibitem{leong2018deep}
Alex~S Leong, Arunselvan Ramaswamy, Daniel~E Quevedo, Holger Karl, and Ling
  Shi,
\newblock ``Deep reinforcement learning for wireless sensor scheduling in
  cyber-physical systems,''
\newblock {\em arXiv preprint arXiv:1809.05149}, 2018.

\bibitem{silva2019optimal}
Vinicius~Lima Silva, Mark Eisen, Konstantinos Gatsis, and Alejandro Ribeiro,
\newblock ``Optimal resource allocation in wireless control systems via deep
  policy gradient,''
\newblock {\em arXiv preprint arXiv:1910.11900}, 2019.

\bibitem{eisen2019control}
Mark Eisen, Mohammad~M Rashid, Konstantinos Gatsis, Dave Cavalcanti, Nageen
  Himayat, and Alejandro Ribeiro,
\newblock ``Control aware radio resource allocation in low latency wireless
  control systems,''
\newblock {\em IEEE Internet of Things Journal}, vol. 6, no. 5, pp. 7878--7890,
  2019.

\bibitem{eisen2019learning}
Mark Eisen, Clark Zhang, Luiz~FO Chamon, Daniel~D Lee, and Alejandro Ribeiro,
\newblock ``Learning optimal resource allocations in wireless systems,''
\newblock {\em IEEE Transactions on Signal Processing}, vol. 67, no. 10, pp.
  2775--2790, 2019.

\bibitem{elgabli2018reinforcement}
Anis Elgabli, Hamza Khan, Mounssif Krouka, and Mehdi Bennis,
\newblock ``Reinforcement learning based scheduling algorithm for optimizing
  age of information in ultra reliable low latency networks,''
\newblock {\em arXiv preprint arXiv:1811.06776}, 2018.

\bibitem{sun2019unsupervised}
Chengjian Sun and Chenyang Yang,
\newblock ``Unsupervised deep learning for ultra-reliable and low-latency
  communications,''
\newblock {\em arXiv preprint arXiv:1905.13014}, 2019.

\bibitem{agiwal2016next}
Mamta Agiwal, Abhishek Roy, and Navrati Saxena,
\newblock ``Next generation 5g wireless networks: A comprehensive survey,''
\newblock {\em IEEE Communications Surveys \& Tutorials}, vol. 18, no. 3, pp.
  1617--1655, 2016.

\bibitem{liu2014ieee}
Jianhan Liu, R~Porat, N~Jindal, et~al.,
\newblock ``Ieee 802.11 ax channel model document,''
\newblock {\em Wireless LANs, Rep. IEEE 802.11--14/0882r3}, 2014.

\bibitem{sutton2000policy}
Richard~S Sutton, David~A McAllester, Satinder~P Singh, and Yishay Mansour,
\newblock ``Policy gradient methods for reinforcement learning with function
  approximation,''
\newblock in {\em Advances in neural information processing systems}, 2000, pp.
  1057--1063.

\end{thebibliography}

\end{document}